\documentclass{article}

\usepackage{arxiv}

\usepackage[utf8]{inputenc} 
\usepackage[T1]{fontenc}    
\usepackage{hyperref}       
\usepackage{cite}
\usepackage{amsmath,amssymb,amsfonts}
\usepackage{graphicx}
\usepackage{textcomp}
\usepackage[table]{xcolor}
\usepackage{longtable}
\usepackage{booktabs}
\usepackage{colortbl}
\usepackage{multirow}
\usepackage{url}
\usepackage{tikz}
\usepackage{pgfplots}
\pgfplotsset{compat=1.17}
\usetikzlibrary{shapes.geometric,arrows.meta,positioning,fit,backgrounds,calc}
\def\BibTeX{{\rm B\kern-.05em{\sc i\kern-.025em b}\kern-.08em T\kern-.1667em\lower.7ex\hbox{E}\kern-.125emX}}

\newcolumntype{L}[1]{>{\raggedright\arraybackslash}p{#1}}
\newcolumntype{Y}{>{\raggedright\arraybackslash}X}

\tikzset{
  box/.style={rectangle, rounded corners=2pt, draw=black!70, line width=0.5pt,
              align=center, inner sep=4pt, font=\footnotesize},
  lane/.style={rectangle, rounded corners=3pt, draw=black!55, line width=0.6pt,
               align=center, inner sep=6pt, font=\small\bfseries},
  cat/.style={rectangle, rounded corners=2pt, draw=black!70, fill=black!6,
              align=center, inner sep=4pt, font=\footnotesize\bfseries},
  leafn/.style={rectangle, rounded corners=1.5pt, draw=black!45, align=left,
               inner sep=3pt, font=\scriptsize},
  flow/.style={-{Stealth[length=2.2mm]}, line width=0.6pt, draw=black!70},
  dflow/.style={-{Stealth[length=2.2mm]}, line width=0.6pt, draw=black!55, dashed},
}

\title{The Rise of AI-Native Software Engineering: Implications for Practice, Education, and the Future Workforce}

\author{
 Mamdouh Alenezi \\
  Saudi Data and Artificial Intelligence (SDAIA)\\
  Riyadh, Saudi Arabia
}

\begin{document}
\maketitle
\begin{abstract}
Generative Artificial Intelligence (GenAI), Large Language Models (LLMs), and emerging Agentic AI constitute the most disruptive transformation in the history of software engineering (SE), reshaping development processes, required competencies, professional roles, and the educational outcomes that universities must deliver. This paper presents a systematic review of 48 verified, influential peer-reviewed publications (2016--2026) drawn from leading venues in software engineering, machine learning, computing education, human--AI collaboration, and software productivity. Studies were discovered, screened, and analyzed through a four-agent research workflow (Literature Discovery, Scientometric Analysis, Curriculum Transformation, and Workforce Impact) and were verified against primary sources. We synthesize the evidence along nine themes and three trajectories---practice, education, and workforce---and report a scientometric inflection in which annual LLM-for-SE output grew roughly five-fold after late 2022. From this synthesis we contribute: (i) a conceptual framework for AI-native software engineering organized around \emph{intent}, \emph{collaboration}, and \emph{verification}; (ii) a nine-dimension competency model spanning specification, critical evaluation, agent orchestration, and metacognition; (iii) a four-phase university curriculum roadmap with AI-resilient assessment; (iv) faculty-development and workforce-transformation strategies; and (v) a prioritized agenda of eleven research gaps. The evidence base is internally contradictory on the magnitude and direction of productivity effects, underscoring that benefits are strongly context-dependent and that educating engineers for judgment, verification, and orchestration---rather than code production alone---is the central challenge of the AI-native era.
\end{abstract}


\section{Introduction}
\label{sec:intro}
Software engineering has evolved through successive waves of methodological and technological change, from structured programming and object-oriented design to agile delivery, DevOps, cloud-native development, and platform engineering. Each of these shifts has altered the instruments of practice, the organization of work, and the competencies expected of practitioners, yet across all of them a stable core endured: human developers authored, reasoned about, and maintained source code, while machines compiled, executed, and tested what people wrote. The emergence of generative artificial intelligence (GenAI), large language models (LLMs), and increasingly agentic systems marks a qualitatively different transition. These technologies do not merely augment individual tasks; they begin to reshape the distribution of cognitive labor across the software development lifecycle, including specification, coding, testing, debugging, documentation, maintenance, and release management. Code-capable assistants such as GitHub Copilot moved from research prototype to mainstream professional practice within roughly two years, and autonomous agents now attempt to resolve real repository issues end to end---reframing the operative question from \emph{``can a model complete a line of code?''} to \emph{``can a system deliver a working, reviewable change?''} Recent studies and roadmapping efforts increasingly characterize this transition as one of the most consequential changes in software engineering in decades \cite{ahmed2025artificial, pezze2025roadmap}.

This transformation is especially significant because it affects software engineering simultaneously as a practice domain, an educational domain, and a labor-market domain. At the level of practice, AI-enabled tools are increasingly capable of generating code, proposing fixes, supporting test creation, and assisting with repository-level problem solving, as illustrated by recent benchmarks and agentic frameworks such as SWE-bench and SWE-agent \cite{jimenez2024swebench, yang2024sweagent}. At the level of education, the same technologies are altering what students are expected to learn, how they are assessed, and which learning outcomes remain meaningful in an AI-mediated environment. At the level of the workforce, they are changing expectations for entry-level competence, professional identity, and the relative value of coding fluency versus judgment, verification, orchestration, and human--AI collaboration. These three dimensions are interdependent: if entry-level coding becomes increasingly automatable, then what universities assess, what employers hire for, and what the professional identity of a ``software engineer'' means must all change together. Any account that treats them separately therefore risks missing the systemic nature of the change.

Despite the rapid diffusion of these tools, the empirical evidence base remains fragmented and, in some respects, contradictory. While some studies report substantial productivity gains, improved developer experience, and accelerated task completion, others find limited benefits, increased overhead, or quality and security trade-offs that depend strongly on task type, developer expertise, and organizational context. The literature also reveals a tension between automation and accountability: the more capable AI systems become at producing code or executing routines, the more important human oversight becomes in validating correctness, robustness, and maintainability. Compounding this, the field is young, fast-moving, and dominated by preprints, so conclusions can shift between the time evidence is generated and the time it is formally reviewed. These inconsistencies make it difficult for educators, curriculum designers, and institutional leaders to draw stable conclusions about what should change in software engineering education and professional preparation.

The implications for higher education are particularly urgent. If the routine production of syntactically correct code becomes increasingly automatable, then software engineering curricula can no longer rely on code generation as the primary proxy for competence. Instead, universities must reconsider the balance among problem framing, computational thinking, specification writing, verification, debugging, ethical reasoning, and collaboration with AI systems. This shift also raises important questions about assessment integrity, academic honesty, and the design of learning experiences that remain robust in an AI-rich environment. At the same time, faculty members require new forms of support to adapt teaching materials, evaluation strategies, and pedagogical models to a rapidly changing technical landscape.

Against this backdrop, the present study systematically synthesizes the evidence from 2016 to 2026 to inform the design of AI-native software engineering education and practice. We ask how GenAI, LLMs, and agentic systems are changing software engineering work; what the empirical evidence indicates regarding productivity, quality, security, and collaboration; how education is responding in terms of pedagogy, tooling, and assessment; how professional roles and skill demands are evolving; and what competency model, curriculum roadmap, and research agenda follow from the accumulated evidence. To address these questions, we analyze a verified corpus of influential peer-reviewed studies and organize the findings into a coherent framework spanning practice, education, and workforce transformation. Concretely, the review is guided by five research questions:

\begin{itemize}
\item \textbf{RQ1.} How have GenAI, LLMs, and agentic systems changed software engineering practice and which tasks are most affected?
\item \textbf{RQ2.} What is the empirical evidence on developer productivity, code quality/security, and human--AI collaboration, and how consistent is it?
\item \textbf{RQ3.} How is computing/SE education responding---pedagogy, tools, and assessment---and what does the evidence show about learning effects?
\item \textbf{RQ4.} How are professional roles, skills demand, and the engineer's identity evolving?
\item \textbf{RQ5.} What competency model, curriculum roadmap, and research agenda follow from the combined evidence?
\end{itemize}

The main contributions of this paper are fivefold. First, we offer a conceptual framework for AI-native software engineering organized around intent, collaboration, and verification. Second, we propose a multi-dimensional competency model that captures the skills needed for effective work in AI-mediated development environments. Third, we present a phased curriculum roadmap that supports educational transformation while preserving academic rigor and assessment validity. Fourth, we derive implications for faculty development and workforce preparation. Fifth, we identify a prioritized agenda of research gaps to guide future empirical and theoretical work. Together, these contributions aim to provide a rigorous foundation for rethinking software engineering for the AI-native era. The remainder of the paper proceeds as follows. Section~\ref{sec:background} situates the work in the history of the discipline and prior reviews; Section~\ref{sec:method} details the review methodology; Sections~\ref{sec:sciento} and \ref{sec:thematic} report the scientometric and thematic findings; Section~\ref{sec:discussion} critically interprets the contradictions in the evidence; Sections~\ref{sec:framework}--\ref{sec:faculty} develop the framework, competency model, curriculum roadmap, and faculty/workforce strategies; and Sections~\ref{sec:agenda}--\ref{sec:data} present the research agenda with threats to validity, the conclusion, and data availability.

\section{Background and Literature Review}
\label{sec:background}

\textbf{Paradigm shifts in software engineering.} The history of software engineering is marked by successive shifts in abstraction, tooling, and the division of labor between humans and machines. Assembly gave way to high-level languages; procedural code gave way to objects and components; waterfall planning gave way to agile and continuous delivery; and on-premise deployment gave way to cloud-native and platform engineering. Each transition raised the level of abstraction at which engineers expressed intent while preserving a deterministic relationship between what was written and what executed. Code-capable LLMs extend this trajectory, but they also introduce a qualitatively different interaction model: the primary interface becomes \emph{natural-language intent}, and the resulting artifact is produced by a stochastic system rather than by a fully deterministic compiler or hand-authored procedure. Hassan et al. frame this transition as ``SE~3.0,'' an intent-first, conversation-driven paradigm in which humans collaborate with AI ``teammates''~\cite{hassan2024se3}. In this view, the central challenge is not only code production, but also the translation, refinement, and verification of intent---a shift that relocates the locus of difficulty from syntax to specification and judgment.

\textbf{Foundations of code LLMs.} The current generation of code-oriented AI systems builds on a sequence of technical milestones. Early foundations include pre-trained representations such as CodeBERT~\cite{feng2020codebert} and CodeT5~\cite{wang2021codet5}, followed by the Codex model and the HumanEval benchmark~\cite{chen2021codex}, competition-level systems such as AlphaCode~\cite{li2022alphacode}, and open, governance-aware models such as StarCoder~\cite{li2023starcoder}. These developments established the feasibility of code generation, completion, and synthesis across a range of programming tasks and progressively expanded the contexts, languages, and problem difficulties that models could address. At the same time, repository-level evaluations such as SWE-bench~\cite{jimenez2024swebench} showed that performance on realistic software engineering problems remains substantially more complex than performance on short, isolated programming tasks. This distinction is important because it highlights the gap between benchmark success on bounded functions and operational usefulness in real development settings, where issues span multiple files, require contextual understanding of a codebase, and demand changes that pass existing test suites.

\textbf{Prior reviews and the gap addressed here.} The literature has expanded rapidly, and several reviews have begun to organize this space. The most comprehensive synthesis to date, Hou et al.~\cite{hou2024llm4se}, reviews 395 LLM-for-SE studies and documents the scale and pace of this growth, mapping models to SE tasks, datasets, and open challenges. In parallel, computing-education syntheses, including the ITiCSE working-group report~\cite{prather2023robotshere} and the \emph{Communications of the ACM} review~\cite{denny2024computing}, examine how educational practice is responding to the same wave of technology. These reviews are valuable but tend to remain within a single literature---either the technical SE literature or the computing-education literature. Our review differs by integrating the \emph{practice}, \emph{education}, and \emph{workforce} literatures in a single framework and by translating the combined evidence into an actionable competency model and curriculum roadmap. This integrative stance is what allows the contradictions observed in one literature (for example, context-dependent productivity effects) to inform recommendations in another (for example, how and when to teach verification and trust calibration).

\section{Methodology}
\label{sec:method}

We adopted a PRISMA-inspired review process executed through a multi-agent research workflow, with the aim of ensuring coverage across the main intellectual streams relevant to AI-native software engineering while preserving traceability, verification, and thematic balance. Four specialized search streams were used to interrogate complementary literatures and reduce the risk of narrow retrieval. \emph{Literature Discovery} targeted code-capable LLMs, agentic software engineering, and SE-task automation. \emph{Scientometric Analysis} focused on publication growth, venues, citation magnitudes, and signals of adoption. \emph{Curriculum Transformation} examined computing-education venues and studies on pedagogical change. \emph{Workforce Impact} addressed productivity, labor economics, and human--AI collaboration. Distributing discovery across four streams was a deliberate design choice: because the phenomenon under study spans technical, educational, organizational, and economic domains, a single search strategy anchored in one community would have systematically under-sampled the others.

\begin{table}[ht]
\centering
\caption{Research questions and principal evidence sources.}
\label{tab:rq}
\renewcommand{\arraystretch}{1.25}
\begin{tabular}{p{0.7cm}p{10.0cm}}
\toprule
\textbf{RQ} & \textbf{Principal evidence themes} \\ \midrule
RQ1 & Foundational models; agentic \& multi-agent SE; SE-task automation \\
RQ2 & Productivity \& human--AI collaboration; quality, security \& trust \\
RQ3 & Computing-education capability/assessment; pedagogy/tools; effects \\
RQ4 & Workforce, labor \& roles; synthesis/vision \\
RQ5 & Cross-thematic synthesis (framework, competencies, roadmap) \\ \bottomrule
\end{tabular}
\end{table}

\textbf{Inclusion and exclusion criteria.} We included peer-reviewed or high-credibility empirical works published between 2016 and 2026 that substantively addressed GenAI, LLMs, or agentic AI in relation to software engineering practice, education, or the workforce. To capture important adjacent evidence, we also included established-laboratory preprints and selected working papers in labor economics when they were credible and directly relevant, since several of the most consequential productivity and adoption results were first disseminated through these channels. We excluded records that were not verifiable, purely promotional material, and studies that were only peripheral to the review focus.

\textbf{Screening and selection.} Records identified by the four streams were screened for topical relevance and methodological credibility. When multiple streams surfaced the same study, duplicates were removed before final selection. The screening process was intentionally iterative: candidate studies were compared across themes, and the corpus was then refined to maintain balance across the three major trajectories of the review---namely practice, education, and workforce---so that no single trajectory dominated the synthesis simply because its literature was larger or faster-growing. From the broader evidence pool, we retained the 48 most influential and representative studies for the final corpus.

\textbf{Verification.} Every retained record was checked against its primary source whenever possible, using authoritative repositories and publishers including arXiv, ACL Anthology, ACM Digital Library, IEEE Xplore, publisher DOI pages, NBER, \emph{Science}, and \emph{Management Science}. A per-record confidence flag was assigned in the companion dataset to document verification status and source reliability. This step was used to improve transparency and to distinguish fully verified records from those requiring more cautious interpretation, which matters in a field where a large share of output circulates as preprints before formal peer review.

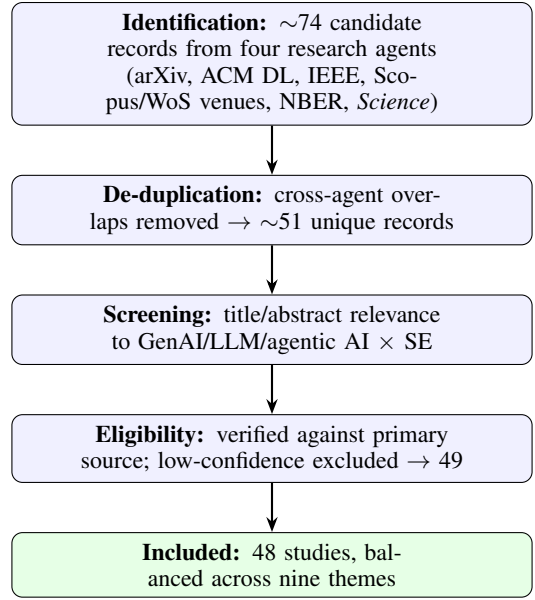
\begin{figure}[t]
\centering
\begin{tikzpicture}[
  node distance=6.5mm and 0mm,
  box/.style={draw, rounded corners, align=center, fill=blue!6, text width=6.6cm, inner sep=4pt, font=\footnotesize},
  arr/.style={-{Stealth[length=2mm]}, thick}]
\node[box] (a) {\textbf{Identification:} $\sim$74 candidate records from four research agents\\(arXiv, ACM DL, IEEE, Scopus/WoS venues, NBER, \emph{Science})};
\node[box, below=of a] (b) {\textbf{De-duplication:} cross-agent overlaps removed $\rightarrow$ $\sim$51 unique records};
\node[box, below=of b] (c) {\textbf{Screening:} title/abstract relevance to GenAI/LLM/agentic AI $\times$ SE};
\node[box, below=of c] (d) {\textbf{Eligibility:} verified against primary source; low-confidence excluded $\rightarrow$ 49};
\node[box, below=of d, fill=green!10] (e) {\textbf{Included:} 48 studies, balanced across nine themes};
\draw[arr] (a)--(b); \draw[arr] (b)--(c); \draw[arr] (c)--(d); \draw[arr] (d)--(e);
\end{tikzpicture}
\caption{PRISMA-style identification, screening, and inclusion of the 48 studies.}
\label{fig:prisma}
\end{figure}

The final corpus was synthesized across nine themes and mapped to the review questions through Table~\ref{tab:rq}, while the complete study list is provided in Table~\ref{tab:corpus}. The resulting dataset supports the PRISMA-style flow reported in Fig.~\ref{fig:prisma} and underpins the conceptual, curricular, and workforce analyses presented in the remainder of the paper.

\section{Scientometric and Temporal Analysis}
\label{sec:sciento}
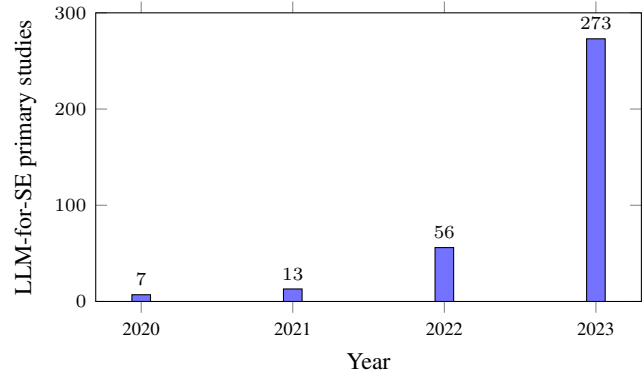
\begin{figure}[t]
\centering
\begin{tikzpicture}
\begin{axis}[
  width=8.8cm, height=5.4cm,
  ybar, bar width=7pt,
  ylabel={\footnotesize LLM-for-SE primary studies},
  xlabel={\footnotesize Year},
  symbolic x coords={2020,2021,2022,2023},
  xtick=data,
  ymin=0, ymax=300,
  nodes near coords, every node near coord/.append style={font=\scriptsize},
  legend style={font=\scriptsize, at={(0.02,0.97)}, anchor=north west},
  tick label style={font=\scriptsize}, ylabel style={font=\footnotesize}]
\addplot[fill=blue!55] coordinates {(2020,7) (2021,13) (2022,56) (2023,273)};
\end{axis}
\end{tikzpicture}
\caption{Field-wide growth of LLM-for-SE research (counts from Hou et al.~\cite{hou2024llm4se}). The $\sim$5$\times$ rise from 2022 to 2023 follows the late-2022 public release of ChatGPT.}
\label{fig:trend}
\end{figure}

The scientometric profile of the corpus reveals a pronounced acceleration in research activity following the emergence of publicly accessible generative AI systems. Consistent with the field-wide analysis reported by Hou et al.~\cite{hou2024llm4se}, the number of primary studies examining LLMs in software engineering increased from 7 in 2020 to 13 in 2021, before rising sharply to 56 in 2022 and 273 in 2023. This approximately five-fold increase coincides with the public release and widespread adoption of ChatGPT, marking a clear inflection point in both research attention and practical deployment (Fig.~\ref{fig:trend}). The growth pattern suggests a rapid transition from exploratory investigations to large-scale scholarly engagement with AI-assisted software development, and it frames the temporal boundary that separates the pre- and post-ChatGPT phases of the literature analyzed here.

A similar trajectory is evident in computing-education research. The ITiCSE Working Group review identified 71 publications addressing generative AI in computing education, with approximately 80\% appearing during the first eight months of 2023 alone~\cite{prather2023robotshere}. This concentration of publications highlights the speed with which educational institutions and researchers responded to the pedagogical implications of generative AI technologies. Collectively, these trends indicate that software engineering and computing education have evolved in parallel, with advances in AI capability driving simultaneous changes in professional practice and instructional design rather than education lagging practice by the years that earlier technological shifts often exhibited.

Evidence from industry surveys further suggests that adoption has progressed alongside capability improvements. Reported usage of AI-assisted development tools increased from approximately 76\% of developers in 2024 to 84\% in 2025. However, this increase in adoption was accompanied by declining levels of self-reported trust in AI-generated outputs. The resulting adoption--trust divergence is noteworthy because it implies that widespread tool utilization does not necessarily correspond to increased confidence in model reliability. This finding reinforces the importance of verification, critical evaluation, and human oversight as core competencies in AI-native software engineering, and it anticipates the ``trust paradox'' developed in the critical discussion.

\begin{figure}[t]
\centering
\begin{tikzpicture}
\begin{axis}[
  width=8.8cm, height=5.0cm,
  xbar, bar width=8pt,
  xlabel={\footnotesize Number of studies in corpus},
  symbolic y coords={Position-vision,Benchmark,Survey,Qualitative-HCI,Experimental,Tool-model-eval},
  ytick=data,
  yticklabels={Position/vision,Benchmark,Survey,Qualitative/HCI,Experimental (RCT/field),Tool/model+eval},
  xmin=0, xmax=16,
  nodes near coords, every node near coord/.append style={font=\scriptsize},
  tick label style={font=\scriptsize}, xlabel style={font=\footnotesize}]
\addplot[fill=orange!70] coordinates {
  (15,Tool-model-eval) (10,Experimental) (8,Qualitative-HCI)
  (5,Survey) (4,Benchmark) (6,Position-vision)};
\end{axis}
\end{tikzpicture}
\caption{Methodological profile of the 48-study corpus (indicative grouping; some studies are mixed-method).}
\label{fig:method}
\end{figure}
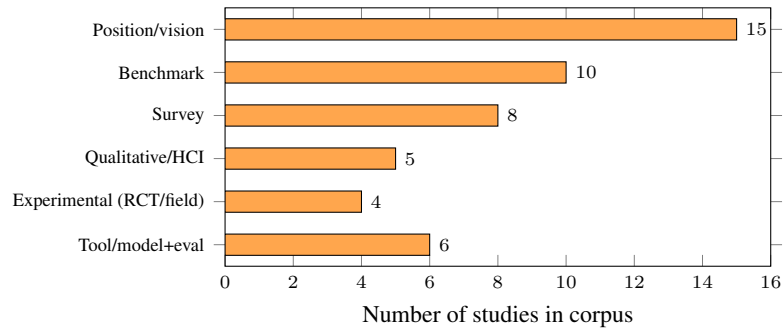

As illustrated in Fig.~\ref{fig:method}, the reviewed corpus exhibits substantial methodological diversity. Studies include tool and model evaluations, controlled and field experiments, qualitative and human--computer interaction investigations, surveys, benchmark studies, and position or vision papers. Such diversity reflects the interdisciplinary nature of the field and the absence of a single dominant methodological paradigm; it also means that synthesizing the evidence requires weighing findings produced under very different epistemic standards, from controlled randomized trials to interpretive qualitative accounts.

The literature is distributed across leading venues in multiple disciplines. Software engineering contributions appear in ICSE, FSE, ASE, TSE, and TOSEM; machine-learning research is represented by ICLR, NeurIPS, ACL, and EMNLP; human--computer interaction studies appear in CHI, OOPSLA, and TOCHI; security-related work is published in S\&P and CCS; educational research is drawn from SIGCSE, ITiCSE, ICER, ACE, and Koli Calling; and workforce-oriented evidence includes contributions from \emph{Science}, \emph{Management Science}, and NBER. This venue diversity underscores the broad impact of generative AI across technical, educational, organizational, and economic domains, and it is one reason a multi-stream discovery strategy was necessary to assemble a balanced corpus.

A persistent characteristic of the field is the high proportion of preprints. More than half of the identified LLM-for-SE publications were disseminated initially through preprint channels before formal peer review. While this pattern reflects the exceptional pace of innovation and knowledge diffusion, it also introduces challenges related to replication, validation, and evidence stability. These concerns are revisited in the research-gap agenda and motivate continued emphasis on reproducibility and longitudinal evaluation, particularly where strong claims about productivity or learning effects rest on single studies that have not yet been independently replicated.

\begin{table}[ht]
\centering
\caption{Distribution of the 48 verified studies across nine research themes.}
\label{tab:themes}
\renewcommand{\arraystretch}{1.2}
\begin{tabular}{p{6.9cm}cc}
\toprule
\textbf{Research theme} & \textbf{\#} & \textbf{\%} \\ \midrule
Foundational Models \& Benchmarks & 6 & 12.5 \\
Agentic \& Multi-Agent SE & 5 & 10.4 \\
SE Task Automation & 2 & 4.2 \\
Code Quality, Security \& Trust & 3 & 6.2 \\
Productivity \& Human-AI Collaboration & 7 & 14.6 \\
Workforce, Labor \& Roles & 6 & 12.5 \\
Computing Education: Capability \& Assessment & 6 & 12.5 \\
Computing Education: Pedagogy \& Tools & 6 & 12.5 \\
Computing Education: Effects \& Adaptation & 3 & 6.2 \\
Synthesis, SLR \& Vision & 4 & 8.3 \\
\midrule
\textbf{Total} & \textbf{48} & \textbf{100.0} \\ \bottomrule
\end{tabular}
\end{table}

\begin{longtable}{c p{4.4cm} c p{2.4cm} c p{5.1cm}}
\caption{Review corpus: 48 studies on LLMs in software engineering and computing education (2016--2026), grouped by theme. Methods: RCT=controlled/field experiment; SUR=survey; QUAL=qualitative/observational; BNCH=benchmark/dataset; TOOL=tool/model+evaluation; SLR=systematic review; POS=position/vision.}
\label{tab:corpus} \\
\toprule
\textbf{ID} & \textbf{Study (cite)} & \textbf{Yr} & \textbf{Venue} & \textbf{Method} & \textbf{Core contribution} \\ \midrule
\endfirsthead

\multicolumn{6}{c}{{\bfseries \tablename\ \thetable{} -- continued from previous page}} \\
\toprule
\textbf{ID} & \textbf{Study (cite)} & \textbf{Yr} & \textbf{Venue} & \textbf{Method} & \textbf{Core contribution} \\ \midrule
\endhead

\midrule \multicolumn{6}{r}{{Continued on next page}} \\ \bottomrule
\endfoot

\bottomrule
\endlastfoot

\multicolumn{6}{l}{\cellcolor{black!8}\textbf{Foundational Models \& Benchmarks}} \\
1 & Codex / HumanEval~\cite{chen2021codex} & 2021 & arXiv & BNCH & Solves 28.8\% of HumanEval at pass@1 (70.2\% at pass@100); repeated sampling sharply boosts solve rate. \\
2 & AlphaCode~\cite{li2022alphacode} & 2022 & Science & TOOL & First system to reach $\sim$top 54\% in simulated Codeforces contests, solving competition-level problems. \\
3 & CodeBERT~\cite{feng2020codebert} & 2020 & EMNLP Find. & TOOL & First bimodal Transformer on NL+code; SOTA on code search and code-to-documentation. \\
4 & CodeT5~\cite{wang2021codet5} & 2021 & EMNLP & TOOL & Identifier-aware pre-training unifies understanding and generation across 8 languages. \\
5 & StarCoder~\cite{li2023starcoder} & 2023 & TMLR & TOOL & Open, governance-aware 15.5B model (8K context, infilling) rivals some closed models. \\
6 & SWE-bench~\cite{jimenez2024swebench} & 2024 & ICLR & BNCH & On 2,294 real GitHub issues, best model solved only $\sim$2\%, exposing the toy-vs-real gap. \\
\multicolumn{6}{l}{\cellcolor{black!8}\textbf{Agentic \& Multi-Agent SE}} \\
7 & ChatDev~\cite{qian2024chatdev} & 2024 & ACL & TOOL & Chat-driven company of role agents produces small software quickly via structured dialogue. \\
8 & MetaGPT~\cite{hong2024metagpt} & 2024 & ICLR & TOOL & Encoding SOPs into role agents cuts cascading hallucination and improves coherence. \\
9 & SWE-agent~\cite{yang2024sweagent} & 2024 & NeurIPS & TOOL & Purpose-built agent-computer interfaces raise autonomous issue-resolution on SWE-bench. \\
10 & ReAct~\cite{yao2023react} & 2023 & ICLR & TOOL & Interleaving reasoning with actions improves task success; underpins tool-using agents. \\
11 & Reflexion~\cite{shinn2023reflexion} & 2023 & NeurIPS & TOOL & Verbal self-reflection stored in memory yields gains across trials, including code tasks. \\
\multicolumn{6}{l}{\cellcolor{black!8}\textbf{SE Task Automation}} \\
12 & LLMs for Program Repair~\cite{xia2023apr} & 2023 & ICSE & TOOL & Applying pre-trained LLMs directly (no fix-training data) surpasses prior repair techniques. \\
13 & TestPilot~\cite{schafer2024testpilot} & 2024 & IEEE TSE & TOOL & Zero-shot JS unit tests reach 70.2\% median statement coverage vs.\ 51.3\% for prior SOTA. \\
\multicolumn{6}{l}{\cellcolor{black!8}\textbf{Code Quality, Security \& Trust}} \\
14 & Asleep at the Keyboard?~\cite{pearce2022asleep} & 2022 & IEEE S\&P & TOOL & $\sim$40\% of 1,689 Copilot programs across MITRE Top-25 CWEs were vulnerable. \\
15 & Insecure Code with AI?~\cite{perry2023insecure} & 2023 & ACM CCS & RCT & AI-assisted users wrote less secure code yet rated it more secure (overconfidence). \\
16 & Reading Between the Lines~\cite{mozannar2024reading} & 2024 & ACM CHI & QUAL & CUPS taxonomy: much developer time goes to verifying/editing suggestions. \\
\multicolumn{6}{l}{\cellcolor{black!8}\textbf{Productivity \& Human-AI Collaboration}} \\
17 & Impact of AI on Productivity~\cite{peng2023impact} & 2023 & arXiv (MSR) & RCT & Copilot users finished an HTTP-server task 55.8\% faster; larger gains for novices. \\
18 & Productivity of Code Completion~\cite{ziegler2022productivity} & 2022 & ACM MAPS & SUR & Acceptance rate predicts perceived productivity better than persistence metrics. \\
19 & Measuring Copilot's Impact~\cite{ziegler2024measuring} & 2024 & CACM & SUR & Across 2,000+ developers, higher acceptance correlates with productivity and flow. \\
20 & Grounded Copilot~\cite{barke2023grounded} & 2023 & OOPSLA & QUAL & Bimodal use: ``acceleration'' (knowing what to write) vs.\ ``exploration''. \\
21 & Expectation vs.\ Experience~\cite{vaithilingam2022expectation} & 2022 & ACM CHI EA & TOOL & Copilot did not improve time/success, yet most preferred it as a starting point. \\
22 & Usability of AI Assistants~\cite{liang2024largescale} & 2024 & ICSE & SUR & Adopted to cut keystrokes/recall syntax; abandoned when output is hard to trust. \\
23 & Early-2025 AI on OSS Devs~\cite{becker2025metr} & 2025 & arXiv (METR) & RCT & Experienced devs $\sim$19\% slower on mature repos yet believed they were faster. \\
\multicolumn{6}{l}{\cellcolor{black!8}\textbf{Workforce, Labor \& Roles}} \\
24 & GenAI in High-Skilled Work~\cite{cui2025effects} & 2025 & Mgmt.\ Science & RCT & Across three firms, Copilot raised completed tasks $\sim$26\%; juniors gained most. \\
25 & Generative AI at Work~\cite{brynjolfsson2023genai} & 2023 & QJE & QUAL & AI assistant raised resolutions/hour 14\% overall, 34\% for novices. \\
26 & Productivity Effects of GenAI~\cite{noy2023experimental} & 2023 & Science & RCT & ChatGPT cut task time $\sim$40\% and raised quality $\sim$18\%; lower performers gained most. \\
27 & GenAI and the Nature of Work~\cite{hoffmann2024nature} & 2024 & HBS WP & TOOL & After Copilot, devs shifted toward core coding away from coordination, persisting $\sim$2 yr. \\
28 & Rapid Adoption of GenAI~\cite{bick2024rapid} & 2024 & NBER WP & SUR & By late 2024, $\sim$23\% of employed U.S.\ workers had used GenAI for work. \\
29 & Toward AI-Native SE (SE 3.0)~\cite{hassan2024se3} & 2024 & ACM TOSEM & POS & Vision of intent-centric development with AI ``teammates'' replacing code-centric work. \\
\multicolumn{6}{l}{\cellcolor{black!8}\textbf{Computing Education: Capability \& Assessment}} \\
31 & The Robots Are Coming (CS1)~\cite{finnieansley2022robots} & 2022 & ACE & BNCH & Codex outperformed most students on CS1 exams, ranking in the top quartile. \\
32 & Will This Be on the Exam? (CS2)~\cite{finnieansley2023cs2} & 2023 & ACE & BNCH & Codex scored top-quartile on CS2 exams; capability extends beyond CS1. \\
33 & Benchmarking ChatGPT/GPT-4~\cite{phung2023benchmarking} & 2023 & ICER & BNCH & GPT-4 beats ChatGPT and nears human tutors, but still lags on grading. \\
34 & LLMs on Beginner Help Requests~\cite{hellas2023exploring} & 2023 & ICER & TOOL & Codex/GPT-3.5 often miss or over-report issues; human guardrails still needed. \\
35 & Copilot on Simple Problems~\cite{wermelinger2023using} & 2023 & SIGCSE TS & QUAL & Solves many simple problems and aids explaining/testing, but is inconsistent. \\
36 & Prompt Engineering for CS1~\cite{denny2023conversing} & 2023 & SIGCSE TS & TOOL & Solves $\sim$half of 166 CS1 problems first try, $\sim$60\% of rest after prompt edits. \\
\multicolumn{6}{l}{\cellcolor{black!8}\textbf{Computing Education: Pedagogy \& Tools}} \\
37 & Auto-Generated Exercises~\cite{sarsa2022automatic} & 2022 & ICER & TOOL & Codex generates novel exercises (with tests) and keyword-steerable explanations. \\
38 & LLM Code Explanations in Class~\cite{macneil2023experiences} & 2023 & SIGCSE TS & SUR & Students engaged with embedded explanations; line-by-line vs.\ high-level valued differently. \\
39 & Student vs.\ LLM Explanations~\cite{leinonen2023comparing} & 2023 & ITiCSE & TOOL & LLM explanations rated more accurate and easier to understand than student ones. \\
40 & Prompt Problems (Promptly)~\cite{denny2024prompt} & 2024 & SIGCSE TS & TOOL & New exercise: students craft prompts to make an LLM produce correct code. \\
41 & CodeHelp~\cite{liffiton2023codehelp} & 2023 & Koli Calling & TOOL & Guard-railed help that avoids giving answers; well received over a 12-week course. \\
42 & AI Code Generators for Novices~\cite{kazemitabaar2023studying} & 2023 & CHI & RCT & Codex improved completion and retention for ages 10--17 without harming later manual work. \\
\multicolumn{6}{l}{\cellcolor{black!8}\textbf{Computing Education: Effects \& Adaptation}} \\
43 & ``It Knows What I Want''~\cite{prather2024weird} & 2024 & ACM TOCHI & QUAL & Identifies novice behaviors (``shepherding,'' ``drifting'') and over-reliance risks. \\
44 & The Widening Gap~\cite{prather2024widening} & 2024 & ICER & QUAL & GenAI sped up strong students but gave strugglers an ``illusion of competence''. \\
45 & From ``Ban It'' to ``Resistance''~\cite{lau2023banit} & 2023 & ICER & QUAL & Instructor stances split between bans and integration; many foresee assessment redesign. \\
\multicolumn{6}{l}{\cellcolor{black!8}\textbf{Synthesis, SLR \& Vision}} \\
30 & LLMs for SE: An SLR~\cite{hou2024llm4se} & 2024 & ACM TOSEM & SLR & Reviews 395 studies (2017--2024), mapping LLMs to SE tasks, data, and challenges. \\
46 & Programming Is Hard~\cite{becker2023programming} & 2023 & SIGCSE TS & POS & Maps opportunities (scaffolding) and challenges (integrity, over-reliance, equity). \\
47 & The Robots Are Here~\cite{prather2023robotshere} & 2023 & ITiCSE-WGR & SUR & Synthesizes 71 articles plus surveys into a roadmap for adapting CS education. \\
48 & Computing Education in GenAI Era~\cite{denny2024computing} & 2024 & CACM & POS & Flagship review of how GenAI reshapes tools, integrity, equity, and open questions. \\
\end{longtable}

\section{Thematic Synthesis of the Literature}
\label{sec:thematic}
Table~\ref{tab:themes} summarizes the thematic distribution of the 48 studies across nine research themes, and Table~\ref{tab:corpus} provides the full corpus with per-study venue, method, and core contribution. To convert this thematic catalogue into an interpretable account, we synthesize the evidence under three trajectories that cut across the themes: the advancing capability of the technology itself (practice), its contested effects on productivity and quality, and the response of education and the workforce.

\subsection{Trajectory 1: From Code Completion to Autonomous Engineering (RQ1)}
The capability frontier advanced from token-level completion to repository-scale autonomy in roughly three years. Foundational models established functional code generation~\cite{chen2021codex,feng2020codebert,wang2021codet5,li2022alphacode,li2023starcoder}; SWE-bench reframed evaluation around real GitHub issues and revealed an initially tiny solve rate~\cite{jimenez2024swebench}. The agentic turn followed quickly: reasoning-and-acting and self-reflection paradigms~\cite{yao2023react,shinn2023reflexion} enabled tool-using agents, and agent--computer interfaces~\cite{yang2024sweagent} together with multi-agent frameworks that encode software roles and standard operating procedures~\cite{qian2024chatdev,hong2024metagpt} pushed autonomous resolution rates sharply upward. In parallel, classic SE tasks were re-tooled: LLM-based program repair surpassed prior techniques~\cite{xia2023apr} and LLM test generation exceeded earlier coverage baselines~\cite{schafer2024testpilot}. The throughline is a migration of human effort \emph{up the abstraction stack}: from writing statements to specifying intent, composing agents, and verifying outcomes. Crucially, the rapid improvement on agentic benchmarks did not eliminate the need for human oversight; rather, it relocated that oversight from line-level authorship to the framing, decomposition, and review of larger units of work.

\subsection{Trajectory 2: Productivity, Quality, and the Contradiction (RQ2)}
The productivity evidence is substantial but \emph{not} uniform. Controlled and field experiments report large gains: a 55.8\% speed-up on a bounded task~\cite{peng2023impact} and an $\sim$26\% increase in completed tasks across three firm RCTs with 4{,}867 developers~\cite{cui2025effects}, with novices benefiting most---echoing skill-compression results in adjacent knowledge work~\cite{brynjolfsson2023genai,noy2023experimental}. Telemetry and survey work links suggestion-acceptance to perceived productivity, fulfillment, and flow~\cite{ziegler2022productivity,ziegler2024measuring}. Yet usability studies find no reliable time/success improvement in some settings~\cite{vaithilingam2022expectation}; behavioral modeling reveals large hidden verification and editing costs~\cite{mozannar2024reading}; and a 2025 RCT found experienced developers were $\sim$19\% \emph{slower} with AI on mature codebases while believing themselves faster~\cite{becker2025metr}. Quality and security findings are sobering: roughly 40\% of generated programs in security-sensitive scenarios were vulnerable~\cite{pearce2022asleep}, and users with AI assistance wrote less secure code while feeling more confident~\cite{perry2023insecure}. Qualitative work explains the mechanism: developers operate in distinct ``acceleration'' and ``exploration'' modes~\cite{barke2023grounded}, and adoption hinges on controllability and comprehensibility~\cite{liang2024largescale}. \textbf{The central empirical lesson is that effect size and even sign are moderated by expertise, task novelty, and codebase maturity}---a finding with direct curricular consequences for teaching judgment and verification rather than treating AI assistance as a uniform accelerant.

\subsection{Trajectory 3: The Education and Workforce Response (RQ3, RQ4)}
Education research moved from alarm to redesign. Codex and successors were shown to outperform most students on CS1 and CS2 assessments~\cite{finnieansley2022robots,finnieansley2023cs2}, and AI tutors approach---but do not match---human tutors on several teaching tasks~\cite{phung2023benchmarking}, while remaining unreliable at diagnosing novice bugs~\cite{hellas2023exploring,wermelinger2023using}. Constructive responses followed: prompt-based pedagogy and ``Prompt Problems'' that teach specification~\cite{denny2023conversing,denny2024prompt}; LLM-generated exercises and explanations~\cite{sarsa2022automatic,macneil2023experiences,leinonen2023comparing}; and guardrailed AI tutors~\cite{liffiton2023codehelp}. Learning-effect evidence is mixed and equity-relevant: scaffolded AI access improved novice outcomes and retention in one controlled study~\cite{kazemitabaar2023studying}, but observational and lab studies document over-reliance, new metacognitive difficulties, and a ``widening gap'' between strong and struggling learners~\cite{prather2024weird,prather2024widening}. Faculty intentions span banning to integration~\cite{lau2023banit}, and the field's agenda-setting works call for rapid, systemic adaptation~\cite{becker2023programming,prather2023robotshere,denny2024computing}. On the workforce side, AI reallocates effort toward core coding and away from coordination~\cite{hoffmann2024nature}, diffuses rapidly across the economy~\cite{bick2024rapid}, and is reframing the engineer's identity toward intent specification and orchestration~\cite{hassan2024se3}. Read together, the education and workforce literatures point in the same direction: the capabilities that remain scarce and valuable are those associated with framing problems, evaluating machine output, and integrating it responsibly into larger systems.

\section{Critical Discussion}
\label{sec:discussion}
Three tensions structure the evidence. First, a \textbf{productivity paradox}: aggregate gains coexist with task-level slowdowns and hidden verification costs, so naive ``X\% faster'' claims are misleading without controlling for expertise and context~\cite{peng2023impact,cui2025effects,becker2025metr,mozannar2024reading}. The same technology that lets a novice complete a bounded task far faster can slow an expert working on a mature, high-stakes codebase, because the marginal value of a suggestion depends on how expensive it is to verify relative to writing the code directly. Second, a \textbf{competence paradox}: the same tools that lift novices' immediate output may undermine the deliberate practice through which durable expertise forms, risking an ``illusion of competence'' in which fluent-looking results mask shallow understanding~\cite{prather2024widening,perry2023insecure}. Third, a \textbf{trust paradox}: adoption rises even as trust falls and measured security worsens~\cite{perry2023insecure,pearce2022asleep}, making \emph{calibrated} trust---knowing when to rely on and when to scrutinize AI output---rather than blanket acceptance or blanket rejection the pivotal skill. These tensions are not anomalies to be averaged away; they are structural features of human--AI collaboration that any educational or organizational response must confront directly. They converge on a single educational implication: the scarce, teachable human capability is no longer code production but \emph{judgment}---specifying intent precisely, evaluating AI output critically, and verifying outcomes responsibly. This conclusion motivates the conceptual framework developed next.

\section{Critical Discussion}
\label{sec:discussion}

Three tensions structure the evidence. First, a \textbf{productivity paradox}: aggregate gains coexist with task-level slowdowns and hidden verification costs, so naive ``X\% faster'' claims are misleading without controlling for expertise and context~\cite{peng2023impact,cui2025effects,becker2025metr,mozannar2024reading}. The same technology that lets a novice complete a bounded task far faster can slow an expert working on a mature, high-stakes codebase, because the marginal value of a suggestion depends on how expensive it is to verify relative to writing the code directly. Second, a \textbf{competence paradox}: the same tools that lift novices' immediate output may undermine the deliberate practice through which durable expertise forms, risking an ``illusion of competence'' in which fluent-looking results mask shallow understanding~\cite{prather2024widening,perry2023insecure}. Third, a \textbf{trust paradox}: adoption rises even as trust falls and measured security worsens~\cite{perry2023insecure,pearce2022asleep}, making \emph{calibrated} trust---knowing when to rely on and when to scrutinize AI output---rather than blanket acceptance or blanket rejection the pivotal skill. These tensions are not anomalies to be averaged away; they are structural features of human--AI collaboration that any educational or organizational response must confront directly. They converge on a single educational implication: the scarce, teachable human capability is no longer code production but \emph{judgment}---specifying intent precisely, evaluating AI output critically, and verifying outcomes responsibly. This conclusion motivates the conceptual framework developed next.

A further implication is that the corpus should not be read as evidence for linear substitution, but rather as evidence for reallocation of effort. The reviewed studies suggest that AI changes where cognitive load sits in the development process: less time may be spent on first-draft code, yet more time is often required for prompt refinement, output checking, debugging, and security review. In this sense, productivity gains are contingent on whether the surrounding workflow is designed to absorb the new verification burden efficiently. Where teams already possess strong systems knowledge, established testing practices, and disciplined review habits, AI can compress low-value work and expand higher-value design and integration work; where those supports are weak, the same tools can increase rework and obscure defects. This is why the literature does not support a universal productivity narrative. It instead indicates that AI magnifies existing process quality, rather than replacing it.

The educational corollary is equally important. If students are allowed to externalize too much early cognitive work, then apparent fluency may rise while conceptual retention weakens, creating a misleading signal of mastery. The competence paradox is therefore not simply about cheating or shortcutting, but about the conditions under which learning is still effortful enough to build robust mental models. In that respect, the emerging consensus is not anti-AI; it is pro-judgment. Educational and professional systems should cultivate engineers who can supervise AI responsibly, preserve conceptual depth, and decide when the machine should lead, assist, or be set aside.

\section{Conceptual Framework for AI-Native Software Engineering}
\label{sec:framework}

\begin{figure}[t]
\centering
\begin{tikzpicture}[font=\footnotesize,
  pillar/.style={draw, rounded corners, align=center, text width=2.05cm, minimum height=1.7cm, fill=blue!12},
  base/.style={draw, rounded corners, align=center, fill=gray!12, text width=6.7cm},
  env/.style={draw, rounded corners, align=center, fill=orange!12, text width=6.7cm}]
\node[env] (env) {\textbf{Ethics, Security \& Responsible-Use Envelope}};
\node[pillar, below=4mm of env.south west, anchor=north west] (p1) {\textbf{Intent}\\\scriptsize specification, prompt \& problem engineering};
\node[pillar, right=2mm of p1] (p2) {\textbf{Collaboration}\\\scriptsize human--AI \& agent orchestration};
\node[pillar, right=2mm of p2] (p3) {\textbf{Verification}\\\scriptsize critical eval, testing, trust calibration};
\node[base, below=4mm of p2.south] (base) {\textbf{Durable CS Foundations}\\\scriptsize algorithms, data structures, systems, architecture};
\draw[-{Stealth[length=2mm]}] (p1.south) -- (base.north -| p1.south);
\draw[-{Stealth[length=2mm]}] (p3.south) -- (base.north -| p3.south);
\begin{scope}[on background layer]
\node[draw, rounded corners, fit=(env)(p1)(p2)(p3)(base), inner sep=3mm, fill=blue!2] {};
\end{scope}
\end{tikzpicture}
\caption{Conceptual framework for AI-native software engineering: three pillars (Intent, Collaboration, Verification) on durable CS foundations, within an ethics-and-security envelope.}
\label{fig:framework}
\end{figure}
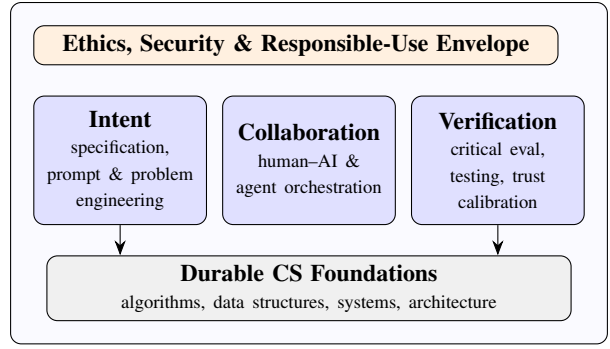

We organize the synthesis into a framework with three interacting pillars---\textbf{Intent}, \textbf{Collaboration}, and \textbf{Verification}---resting on a foundation of durable computer-science fundamentals and bounded by an ethics-and-security envelope (Fig.~\ref{fig:framework}). \emph{Intent} captures the upward migration to specification and prompt engineering: the engineer's primary act becomes expressing what is wanted precisely enough that a stochastic system can act on it~\cite{denny2024prompt,hassan2024se3}. \emph{Collaboration} captures human--AI and human--agent teaming and orchestration, including the composition and supervision of multiple agents that play distinct software roles~\cite{barke2023grounded,qian2024chatdev,yang2024sweagent}. \emph{Verification} captures critical evaluation, testing, security review, and trust calibration---the disciplined scrutiny that converts plausible output into trustworthy software~\cite{perry2023insecure,schafer2024testpilot,mozannar2024reading}. The three pillars are mutually reinforcing: weak intent produces output that is harder to verify, weak verification makes collaboration unsafe, and weak foundations undermine all three by leaving the engineer unable to supervise the system. The framework rests on durable CS foundations because effective oversight of AI output presupposes understanding of algorithms, data structures, systems, and architecture; and it is bounded by an ethics, security, and responsible-use envelope because the documented security and equity risks are not optional add-ons but constraints on every pillar. The framework is the direct source of the competency model (Section~\ref{sec:competency}) and the curriculum roadmap (Section~\ref{sec:curriculum}).

The conceptual value of the framework is that it provides a way to interpret apparently disparate findings within a single structure. Studies of prompting, conversational coding, debugging assistance, agentic orchestration, and security failures all become legible as evidence about different points on the same workflow. The framework makes explicit that AI-native software engineering is not defined by the presence of AI in the toolchain alone, but by a reorganization of responsibility across human and machine actors. Under this interpretation, the engineer does not vanish; instead, the engineer moves upward in abstraction, taking responsibility for defining goals, constraining behavior, and checking outcomes. This shift explains why the literature repeatedly returns to specification, verification, and supervision as central themes, even when the immediate use case appears to be code generation.

The framework also clarifies why durable CS foundations remain indispensable. A supervisor cannot meaningfully evaluate output from a system whose logic, constraints, or failure modes are not understood at least at a working level. Likewise, collaboration with agents becomes unsafe when the human operator lacks the conceptual vocabulary to judge when to intervene. The ethics and security envelope is not a separate layer added after the fact; it is the condition under which all three pillars are acceptable in practice. Taken together, these relations define AI-native SE as a discipline of mediated creation: humans shape intention, coordinate systems of assistance, and retain responsibility for correctness, safety, and accountability.

\section{Competency Model for AI-Native Software Engineering}
\label{sec:competency}

Table~\ref{tab:comp} operationalizes the framework as nine competencies, each mapped to a dominant cognitive level and grounded in corpus evidence. The model deliberately elevates higher-order capabilities---specification, evaluation, orchestration, and metacognition---while retaining foundational CS knowledge as the basis for effective supervision of AI systems. The cognitive-level mapping, expressed in the vocabulary of a revised Bloom's taxonomy, makes explicit why the model is weighted toward \emph{Evaluate} and \emph{Create}: in an environment where generation is cheap, the differentiating human contributions are those that judge, integrate, and direct. Each competency is traceable to specific evidence in the corpus, so the model is not an aspirational wish-list but a synthesis of what the reviewed studies indicate actually distinguishes effective from ineffective work with AI tools. The competencies are intended to be assessable and teachable, and they map onto the curriculum phases in Section~\ref{sec:curriculum}.

The nine competencies also help distinguish surface proficiency from robust capability. Competency C1, for example, is not simply about writing prompts, but about translating problem frames into precise specifications that constrain model behavior. C2 and C3 jointly capture the fact that AI output is only useful when coupled with critical reading, debugging, and verification practices. C4 recognizes that learners and practitioners need reflective habits that prevent overreliance on fluent but incorrect output. C5 reflects the emergence of orchestration as a meaningful engineering skill, particularly as workflows involve multiple tools or agents that must be coordinated rather than used in isolation. C6 serves as the enabling substrate, ensuring that the engineer can reason about the system rather than only interact with it. C7 foregrounds the security and ethical dimension, which is not reducible to compliance but tied directly to safe use. C8 and C9 complete the model by emphasizing communication, collaboration, and adaptability as enduring professional capacities.

Taken together, these competencies define an assessment logic for the AI-native era. The point is not to eliminate traditional measures of correctness, but to supplement them with assessments that reveal whether students or practitioners can inspect output, defend decisions, and operate responsibly in uncertain conditions. This is especially important because fluent interaction with AI can mask weaknesses in explanation, reasoning, or verification. A competency model therefore has value beyond curricular mapping: it gives institutions a language for diagnosing gaps, designing interventions, and tracking progress as AI capabilities continue to evolve.

\begin{table}[ht]
\centering
\caption{AI-native SE competency model (evidence-grounded).}
\label{tab:comp}
\renewcommand{\arraystretch}{1.25}
\footnotesize
\begin{tabular}{p{6.2cm}p{2.4cm}p{2cm}}
\toprule
\textbf{Competency} & \textbf{Cognitive level} & \textbf{Evidence} \\ \midrule
C1 Specification \& intent engineering & Create/Evaluate & \cite{denny2024prompt,denny2023conversing,hassan2024se3} \\
C2 Critical evaluation of AI output & Evaluate/Analyze & \cite{perry2023insecure,pearce2022asleep,mozannar2024reading} \\
C3 AI-assisted debugging \& verification & Apply/Analyze & \cite{schafer2024testpilot,xia2023apr,jimenez2024swebench} \\
C4 Metacognition \& self-regulation & Evaluate & \cite{prather2024widening,prather2024weird,perry2023insecure} \\
C5 Agent orchestration \& tool use & Create/Apply & \cite{yang2024sweagent,hong2024metagpt,qian2024chatdev,yao2023react} \\
C6 Foundational CS \& systems thinking & Understand/Apply & \cite{finnieansley2023cs2,becker2023programming} \\
C7 Security, ethics \& responsible use & Apply/Evaluate & \cite{perry2023insecure,prather2023robotshere,denny2024computing} \\
C8 Human--AI collaboration \& communication & Apply/Create & \cite{barke2023grounded,hoffmann2024nature,hassan2024se3} \\
C9 Continuous learning \& adaptability & Create & \cite{hou2024llm4se,bick2024rapid,prather2023robotshere} \\
\bottomrule
\end{tabular}
\end{table}

\section{Curriculum Implications and University Roadmap}
\label{sec:curriculum}

We propose a four-phase integration model (Table~\ref{tab:road}) that protects deliberate practice early, then progressively shifts toward human--AI teaming and authentic, agentic projects. The unifying principle is \emph{assessment realignment}: because code-writing tasks are now AI-solvable~\cite{finnieansley2022robots,finnieansley2023cs2}, assessment must privilege process, specification, evaluation, and defense of work over artifact production alone. The phasing is itself a response to the competence paradox: early courses deliberately restrict AI on core skill-building so that students develop the mental models that later make them effective supervisors of AI, while later courses progressively open up to AI collaboration as the assessment focus shifts from producing artifacts to directing, evaluating, and defending them. Across all phases, a cross-cutting strand threads responsible use, security, intellectual property, equity, and integrity through the curriculum rather than isolating them in a single ethics course, reflecting the framework's ethics-and-security envelope.

The roadmap is best understood as a gradual transformation of academic judgment rather than a simple increase in tool access. In the foundational phase, the aim is to ensure that students can still reason, trace, and explain computational behavior without relying on generative systems. This is not a retreat from AI, but a necessary condition for later effective use. As learners advance, AI becomes a subject of analysis and a collaborator in bounded tasks, allowing students to experience both the strengths and limitations of model-assisted development. By the time learners reach systems, SE, and capstone contexts, they should be ready to manage larger workflows in which the central challenge is no longer whether AI can produce code, but whether the human team can specify, integrate, audit, and defend what the AI produces.

The roadmap also has institutional significance. It implies that curriculum reform cannot be confined to isolated elective modules or brief policy statements. Programs need a coherent sequence in which learning outcomes, assessment designs, and academic integrity practices reinforce one another. That sequence should be backed by faculty development and shared assessment resources, because otherwise AI-resilient design will remain uneven and difficult to sustain. The roadmap therefore serves both as a pedagogical model and as a governance mechanism for curriculum modernization.

\begin{table}[t]
\centering
\caption{Four-phase AI-native SE curriculum roadmap.}
\label{tab:road}
\renewcommand{\arraystretch}{1.25}
\footnotesize
\begin{tabular}{p{3.3cm}p{4.6cm}p{4.6cm}}
\toprule
\textbf{Phase} & \textbf{Emphasis} & \textbf{AI-resilient assessment} \\ \midrule
1. CS1/CS2 foundations & Durable fundamentals; AI literacy; restrict AI on core skill-building & Invigilated/oral fundamentals; code tracing; prompt-problem tasks \\
2. Core (DS\&A, design) & AI as studied collaborator; design, testing, verification & Test-adequacy tasks; code-review portfolios; design rationales \\
3. SE \& systems & Human--AI teams; agent orchestration; quality \& security at scale & Team projects with AI teammates; defect/security metrics; reflective logs \\
4. Capstone \& electives & Authentic repo-scale, agentic projects; governance & Public defense; contribution \& process evidence \\
\midrule
\multicolumn{3}{p{12.0cm}}{\emph{Cross-cutting:} responsible use, security, IP, equity, and integrity threaded throughout (disclosure statements; trust-calibration exercises).} \\
\bottomrule
\end{tabular}
\end{table}

\section{Faculty Development and Workforce Transformation}
\label{sec:faculty}

\textbf{Faculty development.} Because instructor responses currently range from prohibition to wholesale integration~\cite{lau2023banit}, institutions need structured, evaluated faculty-development programs rather than ad hoc adaptation. We recommend: (i) communities of practice that co-design AI-resilient assessments so that effective designs spread rather than being reinvented course by course; (ii) hands-on training in agentic and tutoring tools with guardrails, building on evidence about where such tools help and where they remain unreliable~\cite{liffiton2023codehelp,phung2023benchmarking}; (iii) shared, openly-licensed assessment banks that lower the cost of moving away from easily-automated artifact production; and (iv) longitudinal evaluation of what faculty actually implement, addressing the well-documented intention--action gap in which stated plans to integrate or restrict AI diverge from classroom practice.

Faculty development should be treated as a design problem, not only a training problem. The central challenge is not merely that instructors need to learn new tools, but that they need new assessment norms, new examples of acceptable use, and new ways of judging student learning when AI is present. Many faculty are likely to benefit from concrete templates that make it easier to ask for evidence of reasoning, reflection, and verification. Others may need support in identifying where AI assistance is pedagogically useful and where it undermines the learning goals of a particular course. A mature faculty-development program therefore combines technical exposure, pedagogical calibration, and policy guidance. It should also create space for instructors to share what works and what fails, because the field is changing too quickly for isolated experimentation to scale reliably.

\textbf{Workforce transformation.} Evidence that AI most benefits novices~\cite{cui2025effects,brynjolfsson2023genai} yet can slow experts on mature systems~\cite{becker2025metr} implies differentiated reskilling rather than a single organization-wide policy. Junior pipelines should emphasize verification and systems understanding to avoid an ``illusion of competence,'' ensuring that early-career engineers build the judgment needed to supervise AI rather than merely accept its output; senior staff, by contrast, need patterns for when \emph{not} to delegate, recognizing the contexts in which manual work remains faster and safer. As effort reallocates from coordination to core coding~\cite{hoffmann2024nature} and roles reorganize around intent and orchestration~\cite{hassan2024se3}, organizations should invest in developer-experience measurement that explicitly accounts for verification overhead~\cite{mozannar2024reading} rather than crediting raw acceptance or output, and in secure-by-default AI workflows given the documented security regressions~\cite{pearce2022asleep,perry2023insecure}. The through-line connecting faculty development and workforce transformation is that both are governance problems as much as training problems: the durable gains come from institutionalizing verification, calibrated trust, and responsible use, not from tool adoption alone.

At the workforce level, the evidence suggests a need for role-sensitive transformation. Entry-level engineers should be equipped to question, inspect, and verify AI-assisted work from the outset, because their value will increasingly lie in disciplined judgment rather than in isolated coding speed. More experienced engineers, meanwhile, must learn to recognize when AI assistance is appropriate, when it adds unnecessary verification burden, and when existing expertise remains the more efficient and safer route. For organizations, this means that talent development should not focus only on accelerating output. It should also aim to preserve organizational memory, strengthen review culture, and prevent overreliance on tools whose performance may vary sharply with task complexity and domain maturity. In that sense, workforce transformation is less about replacing expertise than about redefining what expertise must now include.

\section{Future Research Agenda and Threats to Validity}
\label{sec:agenda}
\textbf{Future research agenda.} Table~\ref{tab:gaps} consolidates eleven gaps and directions synthesized across the corpus. The highest priorities are longitudinal learning and skill-formation studies that follow learners across multiple semesters rather than single sessions; quality- and team-adjusted productivity measurement that captures verification and maintenance costs rather than raw speed; a theory of \emph{when} AI helps versus hinders, expressed in terms of moderators such as expertise, task novelty, and codebase maturity; validated AI-resilient assessment instruments; and equity-focused interventions targeted at the widening gap between strong and struggling learners. These priorities follow directly from the contradictions documented in the critical discussion: each gap marks a place where the existing evidence is either too short-term, too narrowly measured, or too concentrated in a few contexts to support confident generalization.

\begin{table}[ht]
\centering
\caption{Synthesized research gaps and future directions.}
\label{tab:gaps}
\renewcommand{\arraystretch}{1.2}
\footnotesize
\begin{tabular}{p{0.55cm}p{4.2cm}p{5.8cm}}
\toprule
\textbf{ID} & \textbf{Gap} & \textbf{Direction} \\ \midrule
G1 & Longitudinal learning effects & Multi-semester skill-formation studies \\
G2 & Quality-adjusted productivity & Team-level, ecological measurement \\
G3 & Context-dependence of gains & Theory of moderators (expertise, maturity) \\
G4 & Assessment validity \& integrity & Process/oral/specification assessment \\
G5 & Equity \& the widening gap & Adaptive scaffolds; equity metrics \\
G6 & Over-reliance \& metacognition & Trust-calibration pedagogy \\
G7 & Security \& trust at scale & Secure-by-default generation; guardrails \\
G8 & Agentic SE reliability & Verification; human--agent HCI \\
G9 & Role \& identity transformation & Workforce-longitudinal studies \\
G10 & Faculty capacity \& change & Evaluated faculty-development models \\
G11 & Geographic \& open-science gaps & Replication; open benchmarks; collaboration \\
\bottomrule
\end{tabular}
\end{table}

\textbf{Threats to validity.} Several limitations qualify the conclusions drawn here. On \emph{construct and selection validity}, ``influence'' is partly subjective; we mitigated this with multi-agent discovery across complementary literatures and explicit verification, but some relevant work is inevitably omitted. On \emph{currency}, the field moves faster than publication; over half of LLM-for-SE outputs are preprints~\cite{hou2024llm4se}, so some venues and citation magnitudes are reported as orders of magnitude and may have changed since data collection. On \emph{internal validity}, many primary studies use bounded tasks or single institutions; we therefore foreground the \emph{contradictions} in the evidence rather than averaging them away, since a pooled effect size would obscure the very context-dependence that is the central finding. On \emph{external validity}, the education evidence concentrates on introductory programming and a small set of research groups, and output is geographically concentrated; generalization beyond these contexts is uncertain and is itself flagged as a research gap (G11). These threats do not undermine the review's directional conclusions, but they bound the precision with which any single magnitude should be reported or acted upon.

To maintain the same scholarly tone, synthesis-driven style, and conclusion-oriented focus while expanding the section to approximately 240 words, you can use the following version:

\section{Conclusion}
\label{sec:conclusion}

Across 48 verified studies, the evidence is consistent on direction if not magnitude: GenAI, LLMs, and emerging agentic systems are fundamentally reshaping software engineering by shifting human effort away from code authorship and toward intent specification, collaboration, supervision, and verification. Although the literature reports substantial variation in measured outcomes, a common pattern emerges. Productivity gains, learning benefits, and workflow improvements are achievable, but they are neither automatic nor universal. Instead, their realization depends on expertise, task characteristics, organizational context, and the quality of verification practices surrounding AI use.

This review identified three recurring tensions that define the AI-native era: the productivity paradox, the competence paradox, and the trust paradox. Collectively, these tensions suggest that the central challenge is no longer generating software artifacts, but ensuring that humans retain the judgment necessary to direct, evaluate, and govern increasingly capable AI systems. In response, this paper synthesized the evidence into an integrated conceptual framework, a nine-dimension competency model, a four-phase curriculum roadmap with AI-resilient assessment, and a set of faculty-development and workforce-transformation recommendations.

The findings carry important implications for universities, employers, and policymakers. Educational programs that preserve deliberate practice while emphasizing specification, evaluation, verification, and responsible use are likely to produce graduates better prepared for AI-mediated development environments. Likewise, organizations that institutionalize calibrated trust, secure workflows, and effective human oversight will be better positioned to capture the benefits of AI while mitigating its risks. Ultimately, preparing software engineers for the AI-native future is not primarily a technological challenge; it is an educational, organizational, and governance challenge centered on cultivating enduring human judgment in a rapidly evolving technological landscape.

\section{Data Availability}
\label{sec:data}
The collected data is shown within the manuscript in detail as a table. Readers may reach out to the author to obtain a companion spreadsheet that provides full bibliographic and analytical coding for all 48 studies, plus thematic, temporal, methodological, venue, gap, competency, curriculum, and PRISMA sheets.


\bibliographystyle{unsrt}
\bibliography{references}
\end{document}